\documentclass[aps,twocolumn,nofootinbib,groupedaddress,superscriptaddress,longbibliography]{revtex4-2}

\usepackage{amsmath,amssymb}
\usepackage[normalem]{ulem}
\usepackage{graphicx}
\usepackage{multirow,makecell}
\usepackage{bm}
\usepackage{mathtools}
\usepackage{dsfont}
\usepackage{amsfonts}
\usepackage{url}
\usepackage{rotating}
\usepackage{wasysym}
\usepackage{diagbox}
\usepackage{kotex}
\usepackage[dvipsnames]{xcolor}

\usepackage[colorlinks=true,linkcolor=RoyalBlue,citecolor=RoyalBlue,urlcolor=RoyalBlue,hypertexnames=false]{hyperref}

\newcommand{\nocontentsline}[3]{}
\newcommand{\tocless}[2]{\bgroup\let\addcontentsline=\nocontentsline#1{#2}\egroup}

\def\ba#1\ea{\begin{align}#1\end{align}}
\def\bg#1\eg{\begin{gather}#1\end{gather}}
\def\bpm{\begin{pmatrix}}
\def\epm{\end{pmatrix}}

\newcommand{\ourtitle}{
Engineering topological phenomena via field-driven Weyl motion \\
in pyrochlore iridates
}

\allowdisplaybreaks

\begin{document}
\title{\textbf{\ourtitle}}

\author{Jeonghoon \surname{Shin}}
\affiliation{Department of Physics, Soongsil University, Seoul 06978, Korea}

\author{Taekoo \surname{Oh}}
\email{taekoo.oh@ssu.ac.kr}
\affiliation{Department of Physics, Soongsil University, Seoul 06978, Korea}
\affiliation{Origin of Matter and Evolution of Galaxies (OMEG) institute, Soongsil University, Seoul 06978, Korea}

\begin{abstract}
We illustrate the field-driven anomalous Hall engineering via the momentum-space motion of Weyl nodes using pyrochlore iridates under a [111] magnetic field as an example. 
We model the system with a well-established microscopic Hamiltonian and the field effect with the spin rotation from the all-in-all-out to the ferromagnetic state.
Tracking the position of Weyl nodes during the spin rotation, we demonstrate macroscopic anomalous Hall conductivity variations are fundamentally driven by Weyl node motions by 
comparing the full multi-band Kubo formula with the geometric Weyl node separation.
The discrepancy occurs only near a critical spin rotation angle, where several Weyl nodes come into close proximity.
We reveal that the evolution of anomalous Hall conductivity under the magnetic field agrees with reported experiments in an order-of-magnitude and non-monotonicity. 
These results elucidate the field-driven dynamics of momentum-space monopoles and provide a framework for understanding transport in strongly correlated topological materials.
\end{abstract}

\maketitle

\let\oldaddcontentsline\addcontentsline
\renewcommand{\addcontentsline}[3]{}

\section{Introduction}

In condensed matter, electronic correlation and spin-orbit coupling are quintessential energy scales~\cite{witczak2014correlated, schaffer2016recent}.
When these two energy scales are comparable to the electron bandwidth, correlated topological phases, such as quantum spin liquids, topological Mott insulators, and Weyl semimetals, are expected to appear~\cite{wan2011topological, lv2015experimental, weng2015weyl, xu2015discovery}. 
Thus, this regime has attracted considerable attention for decades.

\begin{figure}[t]
    \centering
    \includegraphics[width=1\linewidth]{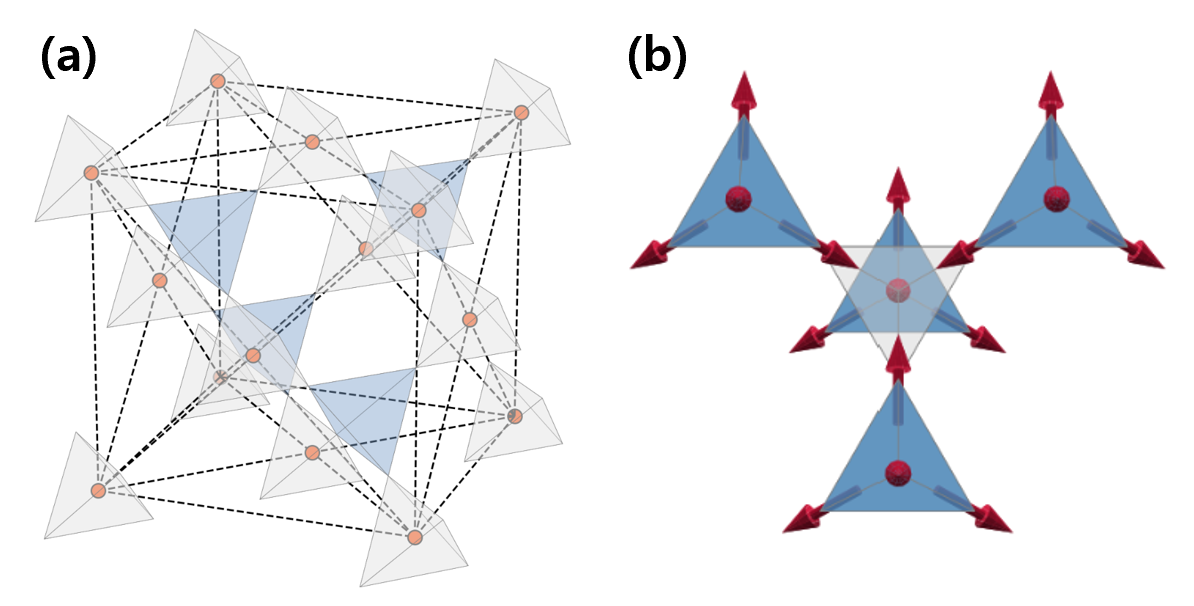}
    \caption{{\bf The basic structure of pyrochlore iridates.} (a) The conventional unit cell of pyrochlore iridates. (b) The all-in-all-out magnetic ordering of pyrochlore iridates.}
    \label{fig:1}
\end{figure}

To this end, pyrochlore iridates (A$_2\text{Ir}_2\text{O}_7$; A is a rare-earth element), whose A and Ir sites form a corner-sharing tetrahedral network, are representative systems exhibiting diverse correlated topological phases under different conditions~\cite{tian2016field,ueda2017magnetic,oh2018magnetic,ueda2018spontaneous,ohtsuki2019strain,li2021correlated,song2022higher,oh2023transverse,shen2024selective}.
In these systems, the low-energy effective physics is governed by the half-filled effective total angular momentum $J_{\text{eff}}=1/2$ states, which are obtained from the localized $5d$ orbitals of Ir$^{4+}$ ions surrounded by an oxygen octahedron [Fig.~\ref{fig:1}(a)]~\cite{oh2018magnetic}.
The effective spin-1/2 moments form the antiferromagnetic all-in-all-out (AIAO) ordering at low temperatures [Fig.~\ref{fig:1}(b)], in which all spins simultaneously point toward or away from the center of each tetrahedron.
Owing to the noncollinear magnetic ordering, time-reversal symmetry is spontaneously broken, driving the system into a magnetic Weyl semimetal (WSM) phase~\cite{wan2011topological, witczak2012topological}.

In a WSM phase, the conduction and valence bands linearly touch at discrete points in momentum space, known as Weyl nodes. 
Each Weyl node possesses a quantized chiral charge and behaves mathematically as a magnetic monopole in three-dimensional $\boldsymbol{k}$-space, an isolated source or sink of Berry curvature flux~\cite{berry1984quantal}. 
This momentum-space Berry curvature acts as an effective fictitious magnetic field that imparts a transverse anomalous velocity to Bloch electrons. 
Consequently, the occupied states produce a macroscopic intrinsic anomalous Hall conductivity (AHC), which manifests the deviation from the linear-B dependence of Hall conductivity and serves as a fundamental transport signature directly reflecting the microscopic topological singularities of the band structure~\cite{nagaosa2010anomalous}.
Although magnetic-field-driven phase transitions in these materials have attracted intense interest, a microscopic understanding of how the engineering of momentum-space singularity motion drives topological transport remains elusive.

In this paper, we illustrate the field-driven engineering of AHC through the motion of Weyl nodes using pyrochlore iridates under a magnetic field along the [111] direction as an example.
First, we model the pyrochlore iridates under the magnetic field with a well-established microscopic Hamiltonian and the spins rotating from the AIAO to the fully aligned ferromagnetic (FM) state as the magnetic field strength increases. 
Next, during the spin rotation, we track the motion of Weyl nodes in the Brillouin zone (BZ) and calculate the intrinsic AHC at the same time. 
The computed intrinsic AHC is around a few tens of S/cm and shows non-monotonic behavior as a function of the rotation angle, in agreement with the experimental results in terms of order of magnitude and non-monotonic dependence on the magnetic field~\cite{ueda2017magnetic}.
We find that the Weyl motion and the AHC are intertwined by comparing the transport properties using two different formulations: the full multi-band Kubo formula and the geometrical Weyl node separation.
The discrepancy only appears near a critical rotation angle, where several Weyl nodes come into close proximity. This proximity makes the band structure near the Weyl nodes flat and close to the Fermi level, so additional Berry curvature contributes to the AHC.

\section{Results}

\subsection{The model}
Pyrochlore iridates are complex materials that exhibit exotic quantum phenomena, which arise from the interplay of strong spin-orbit coupling, electronic correlations, and geometric frustration.
However, as their low-energy physics is determined by the half-filled $J_{\text{eff}} = 1/2$ bands of $5d$ Ir electrons~\cite{oh2018magnetic}, we concentrate on the iridium electrons. 

Therefore, we consider a well-established microscopic Hubbard Hamiltonian on the Ir pyrochlore lattice: $\mathcal{H}=\mathcal{H}_{\text{hopping}} + \mathcal{H}_{\text{on-site}}$~\cite{pesin2010mott,witczak2012topological,witczak2013pyrochlore}. First, we consider hopping up to next-nearest neighbors:
\begin{align}
\mathcal{H}_{\text{hopping}}=&\mathcal{H}_{\text{NN}}+\mathcal{H}_{\text{NNN}}, \label{eq:1}
\end{align}
where
\begin{align}
\mathcal{H}_{\text{NN}}=&\sum_{\langle i,j \rangle} c_{i}^{\dag} (t_1 + i t_2 \boldsymbol{d}_{ij} \cdot \boldsymbol{\sigma}) c_j,\nonumber \\
\mathcal{H}_{\text{NNN}}=& \sum_{\langle\langle i,j \rangle\rangle} c_{i}^{\dag} (t'_{1} + i [t'_{2} \boldsymbol{R}_{ij}\ + t'_{3} \boldsymbol{D}_{ij}] \cdot \boldsymbol{\sigma}) c_j. \nonumber
\end{align}
Here, the hopping amplitudes are given by
\begin{align}
    t_{1}=&\frac{130}{243}t_{oxy}+\frac{17}{324}t_{\sigma}-\frac{79}{243}t_{\pi},\nonumber \\
    t_{2}=&\frac{28}{243}t_{oxy}+\frac{15}{243}t_\sigma-\frac{40}{243}t_{\pi},\nonumber \\
    t'_{1}=&\frac{233}{2916}t'_{\sigma}-\frac{407}{2187}t'_{\pi} \nonumber \\
    t'_{2}=&\frac{1}{1458}t'_{\sigma}+\frac{220}{2187}t'_{\pi}, \nonumber \\
    t'_{3}=&\frac{25}{1458}t'_{\sigma}+\frac{460}{2187}t'_{\pi}. \nonumber
\end{align}
$t_1$ and $t'_1$ are spin-independent hopping amplitudes, which conserve the spin state between nearest neighbors (NNs) and next-nearest neighbors (NNNs), respectively. 
$t_2$ and $t'_2, t'_3$ are spin-dependent hopping amplitudes between NNs and NNNs. 
$t_{oxy}$ is the oxygen-mediated hopping amplitude, while $t_{\sigma}$ and $t_{\pi}$ are direct hopping amplitudes through $\sigma$ and $\pi$ bonds between NNs, respectively. 
$t'_{\sigma}$ and $ t'_{\pi}$ are the hopping amplitudes through $\sigma$ and $\pi$ bonds between NNNs, respectively. 
We set $t'_{\sigma}=0.08t_{\sigma}, t'_{\pi}=0.08t_{\pi}.$
$\boldsymbol{\sigma}=(\sigma_x,\sigma_y,\sigma_z)$ represents the effective spin-1/2 degree of freedom.
\begin{figure}[t]
    \centering
    \includegraphics[width=0.8\linewidth]{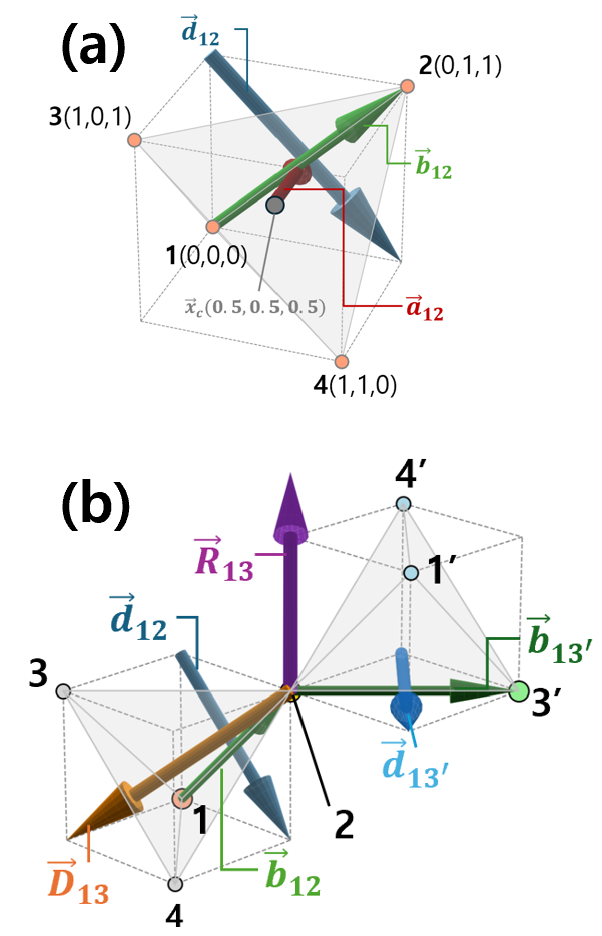}
    \caption{{\bf The parameters used in the model of pyrochlore lattice.}
    (a) The nearest neighbor vectors $\boldsymbol{b}_{12}, \boldsymbol{d}_{12}, \boldsymbol{a}_{12}$ used in Eq.~\ref{eq:1} are shown. (b) The next nearest neighbor vectors $\boldsymbol{R}_{13}, \boldsymbol{D}_{13}$ used in Eq.~\ref{eq:1} are shown.}
    \label{fig:2}
\end{figure}
We choose the sublattice positions of the pyrochlore lattice as shown in Fig.~\ref{fig:2}(a): $\boldsymbol{b}_1=(0,0,0), \boldsymbol{b}_2=(0,1,1), \boldsymbol{b}_3=(1,0,1), \boldsymbol{b}_4=(1,1,0)$. 
In these units, the edge length of the conventional unit cell in Fig.~\ref{fig:1}(a) is $a = 4$. 
$\boldsymbol{b}_{ij} = \boldsymbol{b}_j - \boldsymbol{b}_i$ is the displacement between sublattices.
$\boldsymbol{d}_{ij}$ is the NN Dzyaloshinskii-Moriya (DM)  vector~\cite{elhajal2005ordering, wan2010electronic}, and $\boldsymbol{R}_{ij},\boldsymbol{D}_{ij}$ are the two symmetry-allowed NNN DM vectors. Their definitions are given by
\begin{equation}
\begin{split}
    \boldsymbol{d}_{ij} &= 2\boldsymbol{a}_{ij} \times \boldsymbol{b}_{ij}, \\
    \boldsymbol{R}_{ij} &= \boldsymbol{b}_{ik} \times \boldsymbol{b}_{kj}, \\
    \boldsymbol{D}_{ij} &= \boldsymbol{d}_{ik} \times \boldsymbol{d}_{kj}.
\end{split} \nonumber
\end{equation}
Here, $\boldsymbol{x}_c = (0.5,0.5,0.5)$ is the center of the unit cell, and $\boldsymbol{a}_{ij} = \frac{1}{2}(\boldsymbol{b}_i+\boldsymbol{b}_j) - \boldsymbol{x}_c$ is the displacement of the bond center from the unit cell center. The site $k$ in $\boldsymbol{R}_{ij}$ and $\boldsymbol{D}_{ij}$ indicates the intermediate site between $i$ and $j$. 
In Figs.~\ref{fig:2}(a,b), we depict examples of $\boldsymbol{b}_{ij}, \boldsymbol{d}_{ij}, \boldsymbol{R}_{ij}$, and $\boldsymbol{D}_{ij}$.

On the other hand, the Hubbard repulsion term is given by:
\begin{equation}
\mathcal{H}_{\text{on-site}} = U \sum_{a} \hat{n}_{a,\uparrow} \hat{n}_{a,\downarrow},
\end{equation}
where $U$ is the on-site Coulomb energy, and $\hat{n}_{a,\alpha}$ is the number operator for electrons with spin $\alpha$ at sublattice $a$. Under the mean-field approximation at half-filling, this interacting Hamiltonian is reduced to a single-particle magnetic exchange term:
\begin{equation}
\mathcal{H}_{\text{on-site}}^{\text{MF}} = -\frac{U}{2} \sum_{a} \boldsymbol{m}_a \cdot \boldsymbol{\sigma} + const.,
\end{equation}
where $\boldsymbol{m}_a$ is the local magnetization vector determined by the spin expectation value at sublattice $a$. This term acts as a local Zeeman exchange field that drives the topological phase transitions~\cite{oh2018magnetic}. 
Below, we use $\mathcal{H}^{\text{MF}} = \mathcal{H}_{\text{hopping}} + \mathcal{H}_{\text{on-site}}^{\text{MF}}$ instead of $\mathcal{H}$ for computations. 
We provide further computational details in the Supplemental Materials (SM).
\begin{figure}[t]
    \centering
    \includegraphics[width=1\linewidth]{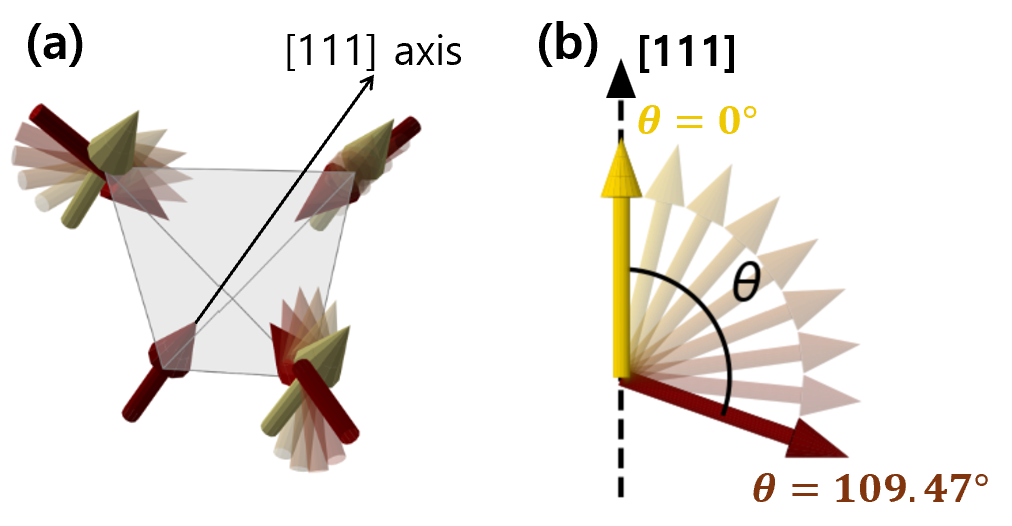}
    \caption{{\bf The assumption on the effect of magnetic field.} 
    (a) When a magnetic field is applied to [111] direction, we assume that the spins gradually rotate from AIAO to FM state. As one spin is already parallel to [111], other three spins rotate toward [111] direction. 
    (b) The initial position of three spins are represented by the rotation angle $\theta =109.47^\circ$ and their final position are at $\theta = 0^\circ$.}
    \label{fig:3}
\end{figure}

To simulate the effect of an external magnetic field along [111] on the local spins, we assume that the magnetic field rotates the spins gradually toward the [111] direction without changing their magnitudes~\cite{ueda2017magnetic}. [See Fig.~\ref{fig:3}(a).]
We begin from the all-in state, which is preferred to the all-out state under a field along the +[111] direction~\cite{kim2018unconventional,kim2020strain}, and rotate the spins toward the FM state where all spins are parallel to the [111] direction.
Since one of the four spins in the unit cell is already parallel to the [111] direction in the all-in state, the spin rotation occurs for the other three spins.

We use the canting angle $\theta$ as a phenomenological parameter for field-induced spin reorientation while keeping the local moment magnitude fixed. Along the assumed canting path, increasing the field along [111] is represented by decreasing $\theta$. A quantitative relation between $\theta$ and the applied field is not determined in the present model. Our comparison with experiment therefore concerns the order of magnitude and the overall non-monotonic trend of the AHC.

\begin{figure}[t]
    \centering
    \includegraphics[width=0.8\linewidth]{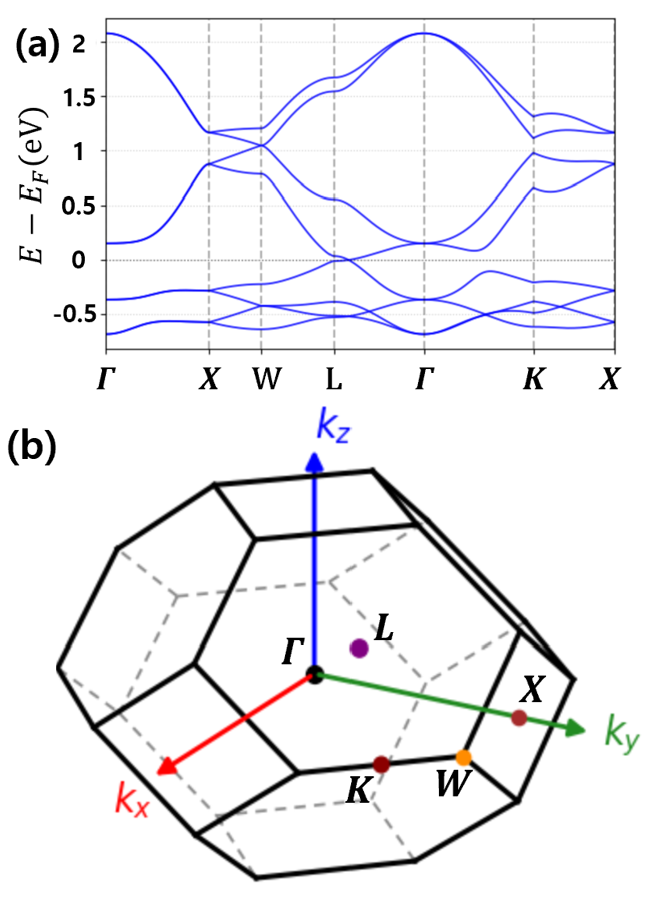}
    \caption{{\bf Weyl semimetal phase of pyrochlore iridates at AIAO state.} (a) The band structure in the condition $U=1.485, t_{\sigma}=-0.8, m=0.35$ at $\theta = 109.47^\circ$. 
    This is a WSM with four pairs of Weyl nodes, which reside at the $\Gamma-L$ line. 
    (b) The first Brillouin zone of pyrochlore lattice (fcc).}
    \label{fig:4}
\end{figure}

Hence, we can express the magnetization vectors of the three spins as $\boldsymbol{m}_a = m \hat{\boldsymbol{n}}_a(\theta)$, where $\hat{\boldsymbol{n}}_a(\theta)$ is the unit vector indicating the spin direction continuously parameterized by the spin canting angle $\theta$. 
$\theta$ denotes the angle between the direction of each spin and the [111] direction, which is assumed to be the same for the three rotating spins due to the remaining threefold rotation symmetry. 
$\theta$ continuously varies from $109.47^{\circ}$ (AIAO) to $0^{\circ}$ (FM).
Then, $\hat{\boldsymbol{n}}_a$ can be described by
\begin{equation}
\begin{split}
&\hat{\boldsymbol{n}}_1 = \hat{\boldsymbol{n}}_{1,\text{ini}},\\
&\hat{\boldsymbol{n}}_2 = \hat{\boldsymbol{n}}_{1,\text{ini}}\cos\theta + \hat{\boldsymbol{n}}_{2,\text{ini}}\sin\theta,\\
&\hat{\boldsymbol{n}}_3 = \hat{\boldsymbol{n}}_{1,\text{ini}}\cos\theta + \hat{\boldsymbol{n}}_{3,\text{ini}}\sin\theta,\\
&\hat{\boldsymbol{n}}_4 = \hat{\boldsymbol{n}}_{1,\text{ini}}\cos\theta + \hat{\boldsymbol{n}}_{4,\text{ini}}\sin\theta,\\
\end{split}
\end{equation}
with
\begin{equation}
\begin{split}
&\hat{\boldsymbol{n}}_{1,\text{ini}} = (1,1,1)/\sqrt{3},\\
&\hat{\boldsymbol{n}}_{2,\text{ini}} = (2,-1,-1)/\sqrt{6},\\
&\hat{\boldsymbol{n}}_{3,\text{ini}} = (-1,2,-1)/\sqrt{6},\\
&\hat{\boldsymbol{n}}_{4,\text{ini}} = (-1,-1,2)/\sqrt{6}.
\end{split} \nonumber
\end{equation}
We place the spins at the positions $\boldsymbol{b_1}$ to $\boldsymbol{b_4}$ defined above.

We show the energy band structure of the system in the AIAO state in Fig.~\ref{fig:4}(a).
We use the parameters $U=1.485$, $t_\sigma=-0.8$, and $m=0.35$, which are chosen to realize a Weyl semimetal based on previous reports~\cite{witczak2013pyrochlore,oh2018magnetic,kim2020strain}.
There are a total of four pairs of Weyl nodes on the $\Gamma-L$ lines.

\subsection{Magnetic-field-driven Weyl motion and AHC}

To evaluate the intrinsic AHC and investigate its origin, we employ two complementary approaches. 
The first method is based on linear response theory using the Kubo formula. 
According to the Kubo formula, the intrinsic AHC is determined by integrating the Berry curvature of the occupied bands over the BZ. 
The conductivity tensor $\sigma_{xy}$ is given by
\begin{align}
\begin{split}
\sigma_{xy}^{\text{Kubo}}=&-\frac{e^2}{\hbar}\int_{BZ}\frac{d^3\boldsymbol{k}}{(2\pi)^3}\sum_n f_n(\boldsymbol{k})\Omega_{n}^{z}(\boldsymbol{k}),
\\
\Omega_{n}^{z}(\boldsymbol{k})=&-2\hbar^2 \sum_{m\neq n}\text{Im}[\frac{\langle n|v_x(\boldsymbol{k})|m  \rangle \langle m|v_y(\boldsymbol{k})|n \rangle}{(E_n(\boldsymbol{k})-E_m(\boldsymbol{k}))^2}],
\end{split}
\end{align}
Here, $v_x(\boldsymbol{k}), v_y(\boldsymbol{k})$ are the velocity operators, $E_n(\boldsymbol{k})$ is the eigenenergy, and $f_n(\boldsymbol{k})$ is the Fermi-Dirac distribution function.

The second method relies on the geometric distribution of topological singularities in $k$-space, which provides a more intuitive understanding of the Weyl semimetal phase. 
When spins rotate, three pairs of Weyl nodes deviate from the $\Gamma-L$ lines due to crystalline symmetry breaking. 
Thus, we systematically scan the BZ to pinpoint the exact $k$-space coordinates and the chiral charge of Weyl nodes. [See SM for details.]
Note that Weyl nodes should exist in pairs with opposite chiralities according to the Nielsen-Ninomiya theorem.
Then, the intrinsic AHC is directly proportional to the momentum-space separation $\Delta \boldsymbol{K}$ between such a pair, if Weyl nodes are exactly at the Fermi level in two-band systems~\cite{burkov2011weyl}. 
The conductivity can be expressed as:
\begin{equation}
\sigma_{xy}^{\text{node}}=\frac{e^2}{2\pi h}\Delta K_z, \label{eq:6}
\end{equation}
where $\Delta K_z$ is the z-component of the summed node separation vector or Chern vector throughout the first BZ.
Note that for both methods, due to the threefold rotation symmetry of the system, $\sigma_{xy} = \sigma_{yz} = \sigma_{zx}$.

\begin{figure}
    \centering
    \includegraphics[width=1\linewidth]{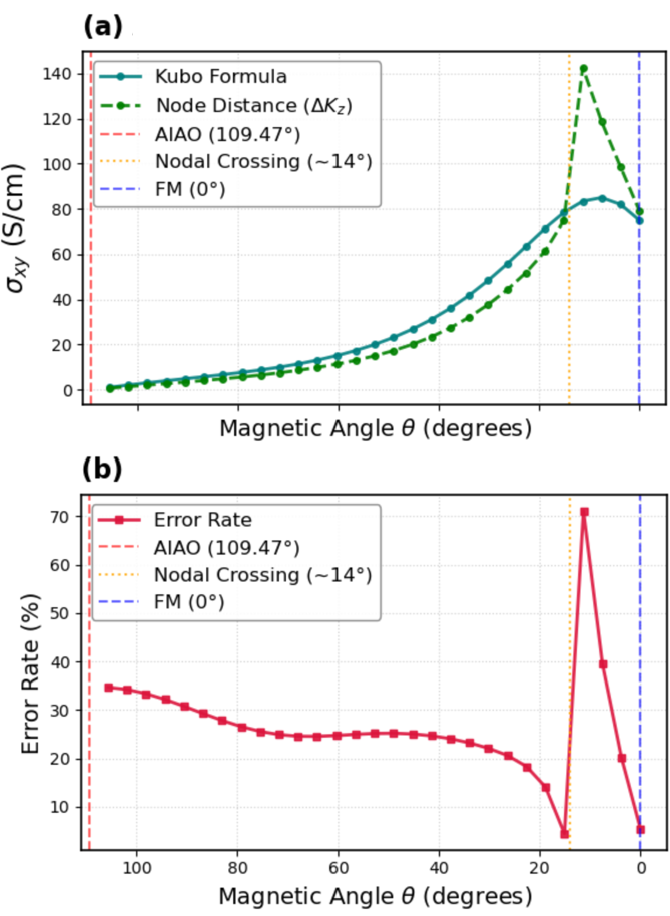}
    \caption{{\bf The computed AHC as a function of the spin rotation angle $\theta$.}
    (a) The comparation between the Kubo formula with node distance method.
    The blue solid line is obtained by the Kubo formula, while the green dotted line is computed by the Weyl node separation. 
    (b) The error rate of AHC between $\sigma^{Kubo}_{xy}$ and $\sigma^{node}_{xy}$ at each $\theta$. The red solid line is the error rate calculated by formula ($\sigma^{Kubo}_{xy}$ - $\sigma^{node}_{xy}$)/$\sigma^{Kubo}_{xy}\times100~ \%$.
    Yellow vertical dotted line at $\theta_c \approx 14^\circ$ is the critical angle where two methods begin to differ. The difference appear when several Weyl nodes come into close proximity, as shown in Fig.~\ref{fig:6}.}
    \label{fig:5}
\end{figure}

The results are summarized in Fig.~\ref{fig:5}.
From these results, we make the following observations.
First, in Fig.~\ref{fig:5} (a), we confirm that the AHC from the Weyl node separation method (dotted line) is close to that from the Kubo formula method (solid line) for a wide range of $\theta$. The error rate described in Fig.~\ref{fig:5}(b), shows that from 65 to 90 percent of $\sigma_{xy}^{\text{Kubo}}$ could be explained by $\sigma_{xy}^{\text{node}}$ above the critical angle $\theta_c \approx 14^\circ$. 
Here, the error in this regime can be come from the shift of node energy during the spin rotation.

The critical angle at which the two methods begin to severely differ is $\theta_c \approx 14^\circ$ (yellow dotted line).
Less than 30 percent of $\sigma_{xy}^{\text{Kubo}}$  could be explained. 
This indicates that the intrinsic AHC originates mainly from the redistribution of Weyl nodes in the BZ due to the spin rotation.
Second, the experimental results show a Hall conductivity of about a few tens of S/cm, and reveal that the magnetic field dependence of Hall conductivity is non-monotonic; the Hall conductivity increases at a low field and peaks at a high field.
Since one can consider a decrease in $\theta$ as an increase in magnetic field strength, the computed results agree with the experimental results~\cite{ueda2017magnetic} in terms of order of magnitude and non-monotonicity.

\begin{figure}[t]
    \centering
    \includegraphics[width=1\linewidth]{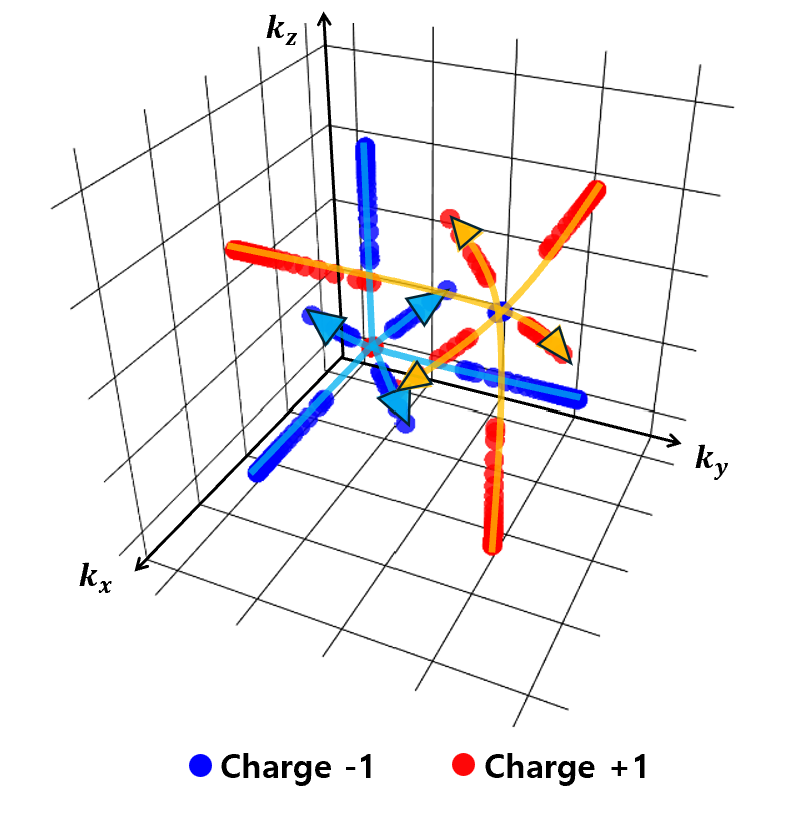}
    \caption{{\bf The motion of Weyl nodes due to the spin rotation.}
    Red and blue circles are Weyl node which has $\pm1$ topological charge. 
    As $\theta$ changes from $109.47^{\circ}$ to $0^\circ$, each nodes move along the arrows.}
    \label{fig:6}
\end{figure}

We further investigate why the two methods give different results at $\theta \lesssim 14^\circ$.
Figure~\ref{fig:6} shows the motion of the four Weyl nodes due to the spin rotation. 
The starting points of the arrows denote $\theta=109.47^\circ$ (AIAO) and the end points denote $\theta=0^\circ$ (FM). 
In the AIAO state, there are four pairs of Weyl nodes on the $\Gamma-L$ lines, since the magnetic point group of the primitive unit cell (tetrahedron) is $-4'3m'$. 
This magnetic point group includes threefold unitary rotations, twofold unitary rotations, fourfold antiunitary rotoinversions, and antiunitary mirrors.
The axes of the threefold rotations are the $\Gamma-L$ lines, so the Weyl nodes cannot leave these lines as long as AIAO order persists.

However, as $\theta$ goes to zero, three nodes whose topological charge is $\pm$1 deviate from the $\Gamma-L$ lines and move toward another node whose topological charge is $\mp$1. 
They cross near $\theta \approx 14^\circ$ and follow their own curved paths in the mirror plane.
The spin rotation reduces the magnetic point group to $3m'$, which includes threefold rotations around the $[111]$-axis and antiunitary mirrors whose planes are parallel to $[111]$. 
Therefore, while the pair of Weyl nodes on the [111] $\Gamma-L$ line remains on this line due to the threefold rotation, the other three pairs of Weyl nodes deviate from the other $\Gamma-L$ lines but reside in the antiunitary mirror planes.

Near $\theta \approx 14^\circ$, the distances between different Weyl nodes are infinitesimal, so the band structure nearby Weyl nodes become flat and close to the Fermi level.
Therefore, Berry curvature comes from different parts of the bands instead of Weyl nodes, and the computed Berry curvature deviate away from the Weyl node separation. 

\section{Discussion}

Here, we further discuss three points.
First, we address the stability of the double WSM phase.
At the exact crossing of the node trajectories in Fig.~\ref{fig:6}, three $C=\pm1$ Weyl nodes and one $C=\mp1$ Weyl node merge and form a double Weyl node with chiral charge $\pm 2$.
Then, only a pair of double Weyl nodes remains in the BZ, so the double WSM phase is achieved.
However, the double WSM phase is unstable in that it appears only at a particular value of $\theta$. 

Next, we address how the spin canting determines the trajectories of Weyl nodes. When there is no magnetic ordering, the low-energy band is described by the quadratic band crossing (QBC) at the $\Gamma$ point due to the high crystalline symmetry of the system~\cite{kondo2015quadratic}.
When the magnetic ordering develops, the spins residing in the primitive unit cell collectively form a cluster multipole~\cite{suzuki2017cluster}.
Since the wavefunction in the QBC also carries the cluster multipole, the spin ordering can couple to the QBC and break it into pairs of Weyl nodes~\cite{oh2018magnetic}. 
AIAO order carries an $A_2$-octupole, FM order carries a $T_1$-dipole, and the intermediate state carries an $A_2$-octupole, a $T_1$-dipole, and a $T_1$-octupole~\cite{kim2020strain}.
The continuous change of the multipole during the spin rotation determines the type of topological semimetal and the trajectories of the Weyl nodes.

We emphasize the difference of our work with the previous studies~\cite{oh2018magnetic,kim2020strain}. 
Previous studies have established magnetic control of topological semimetal phases and the role of cluster multipoles in anomalous Hall transport in pyrochlore iridates.
Here, we track the continuous evolution of Weyl-node trajectories and the intrinsic AHC along a prescribed spin-canting path within the same microscopic model. 
By directly comparing the full multiband Kubo response with the node-separation estimate, we assess how well the evolving Weyl-node configuration accounts for the Hall response.

In summary, we systematically investigate the engineering of topological phenomena via the field-driven motion of Weyl nodes using pyrochlore iridates under a [111]-oriented external magnetic field as an example. 
By employing a well-established microscopic Hamiltonian, we continuously track the trajectory of Weyl nodes in momentum space and the change in AHC as the spin canting angle $\theta$ is varied from the AIAO ($\theta=109.47^\circ$) state to the fully aligned FM state ($\theta=0^\circ$). 
We successfully link the continuous evolution of the macroscopic AHC to the field-driven Weyl motion.
By illustrating that the field-driven dynamics of momentum-space monopoles can be used to engineer topological properties, we emphasize the transport phenomena of correlated materials and point to the potential application of engineered topological materials to condensed matter devices.

\begin{acknowledgments}
T.O. was supported by the Pioneer Project for Future-Oriented Convergence Technology (Challenge Type; RS-2024-00416976) and Basic Science Research Program (RS-2025-16065011) through the National Research Foundation of Korea (NRF) funded by the Ministry of Science and ICT of Korean government, Basic Science Research Program (RS-2026-25470523) through the National Research Foundation of Korea (NRF) funded by the Ministry of Science and ICT of Korean government.
T.O. and J.S. were supported by G-LAMP (RS-2025-25441317) through the National Research Foundation of Korea (NRF) funded by the Ministry of Education of Korean government.
This work was also supported by the Nano \& Material Technology Development Program through the NRF, funded by the Ministry of Science and ICT (RS-2026-25607606).
\end{acknowledgments}

\bibliography{Refs.bib}

\let\addcontentsline\oldaddcontentsline

\clearpage
\onecolumngrid
\begin{center}
\textbf{\large Supplemental Materials for ``Engineering topological phenomena via field-driven Weyl motion in pyrochlore iridates"}
\end{center}

\setcounter{section}{0}
\setcounter{figure}{0}
\setcounter{equation}{0}
\setcounter{table}{0}
\renewcommand{\thefigure}{S\arabic{figure}}
\renewcommand{\theequation}{S\arabic{equation}}
\renewcommand{\thesection}{S\arabic{section}}
\renewcommand{\thetable}{S\arabic{table}}
\tableofcontents
\hfill \\
\onecolumngrid

\section*{Supplementary Note 1: Construction of the Momentum-Space Tight-Binding Hamiltonian}

In the main text, the tight-binding Hamiltonian is presented in its real-space second-quantization representation to provide an intuitive understanding of the physical interactions:
\begin{equation}
\mathcal{H}_{\text{hopping}} = \sum_{\langle i,j \rangle} c_{i}^{\dag} (t_1 + i t_2 \boldsymbol{d}_{ij} \cdot \boldsymbol{\sigma}) c_j + \sum_{\langle\langle i,j \rangle\rangle} c_{i}^{\dag} (t'_{1} + i [t'_{2} \boldsymbol{R}_{ij} + t'_{3} \boldsymbol{D}_{ij}] \cdot \boldsymbol{\sigma}) c_j,
\end{equation}
where $c_i^\dag$ ($c_i$) is the electron creation (annihilation) operator at site $i$, and $\boldsymbol{\sigma} = (\sigma_x, \sigma_y, \sigma_z)$ is the vector of Pauli matrices.

For numerical calculations of the band structure and topological properties (such as the Berry curvature and anomalous Hall conductivity), we perform a Fourier transform to obtain the $8 \times 8$ Bloch Hamiltonian $H(\boldsymbol{k})$ in momentum space. The basis is chosen as $\Psi_{\boldsymbol{k}} = (c_{0\uparrow}, c_{0\downarrow}, c_{1\uparrow}, c_{1\downarrow}, c_{2\uparrow}, c_{2\downarrow}, c_{3\uparrow}, c_{3\downarrow})^T$, where the indices $0, 1, 2, 3$ denote the four Ir sublattices within the primitive unit cell of the pyrochlore lattice. The numerical implementation maps the real-space equation to the $\boldsymbol{k}$-space matrix elements $H^{ab}(\boldsymbol{k})$ (where $a, b \in \{0, 1, 2, 3\}$) as follows:

\subsection*{1. Magnetic Exchange (Mean-Field) Term}
Before considering the hopping terms, the local magnetic exchange interaction driven by the on-site Hubbard $U$ (derived rigorously in Supplementary Note 2 as $\mathcal{H}_{\text{on-site}}^{\text{MF}}$) is implemented as a block-diagonal matrix. The $a$-th diagonal block corresponding to sublattice $a$ is given by:
\begin{equation}
\mathcal{H}_{\text{mag}}^{aa} = - \frac{Um}{2} \hat{\boldsymbol{n}}_a(\theta) \cdot \boldsymbol{\sigma},
\end{equation}
where $m$ is the uniform magnitude of the local magnetic moment, and $\hat{\boldsymbol{n}}_a(\theta)$ is the unit vector indicating the spin quantization direction. This direction is smoothly parameterized by the canting angle $\theta$ from the All-In-All-Out (AIAO, $\theta=109.47^\circ$) state to the ferromagnetic (FM, $\theta=0^\circ$) state along the $[111]$ direction. This directly corresponds to the $-(U * m_{val} / 2.0) * m_a\cdot \boldsymbol{\sigma}$ term in our numerical code.

\subsection*{2. Nearest-Neighbor (NN) Hopping}
The pyrochlore lattice consists of a face-centered cubic (FCC) Bravais lattice with a four-atom basis. Let $\boldsymbol{b}_a$ be the position of sublattice $a$. The vector connecting two nearest-neighbor sublattices is defined as $\boldsymbol{b}_{ab} = \boldsymbol{b}_b - \boldsymbol{b}_a$. By applying the Fourier transform and grouping the bonds related by spatial inversion symmetry ($\boldsymbol{b}_{ab}$ and $-\boldsymbol{b}_{ab}$), the exponential phase factors $e^{i \boldsymbol{k} \cdot \boldsymbol{b}_{ab}} + e^{-i \boldsymbol{k} \cdot \boldsymbol{b}_{ab}}$ reduce to a cosine function. The NN block matrix element between sublattice $a$ and $b$ ($a \neq b$) becomes:
\begin{equation}
H_{\text{NN}}^{ab}(\boldsymbol{k}) = 2 \left( t_1 \sigma_0 + i t_2 \boldsymbol{d}_{ab} \cdot \boldsymbol{\sigma} \right) \cos(\boldsymbol{k} \cdot \boldsymbol{b}_{ab}).
\end{equation}
This is mapped exactly to the \texttt{term\_NN} in the code, where $\boldsymbol{d}_{ab}$ is the symmetry-allowed Dzyaloshinskii-Moriya (DM) vector for the nearest-neighbor bond.

\subsection*{3. Next-Nearest-Neighbor (NNN) Hopping}
On the pyrochlore lattice, two sublattices $a$ and $b$ are next-nearest-neighbors if they are connected through an intermediate nearest-neighbor site $c$ (where $c \neq a$ and $c \neq b$). To systematically sum over all NNN pathways in the unit cell, our algorithm iterates over this intermediate mediating site $c$.

For a specific path $a \rightarrow c \rightarrow b$, the geometric vectors $\boldsymbol{R}_{ab}^{(c)}$ and $\boldsymbol{D}_{ab}^{(c)}$ in the main text are explicitly constructed using cross products of the nearest-neighbor vectors:
\begin{align}
\boldsymbol{R}_{ab}^{(c)} &= \boldsymbol{b}_{ac} \times \boldsymbol{b}_{cb}, \\
\boldsymbol{D}_{ab}^{(c)} &= \boldsymbol{d}_{ac} \times \boldsymbol{d}_{cb}.
\end{align}
Letting the relative momentum phase for this path be $\boldsymbol{k} \cdot (\boldsymbol{b}_{cb} - \boldsymbol{b}_{ac})$, we obtain the NNN block matrix:
\begin{equation}
H_{\text{NNN}}^{ab}(\boldsymbol{k}) = 2 \sum_{c \neq a,b} \left( t'_1 \sigma_0 (1-\delta_{ab}) + i \left[ t'_2 \boldsymbol{R}_{ab}^{(c)} + t'_3 \boldsymbol{D}_{ab}^{(c)} \right] \cdot \boldsymbol{\sigma} \right) \cos(\boldsymbol{k} \cdot (\boldsymbol{b}_{cb} - \boldsymbol{b}_{ac})).
\end{equation}
This geometric structural expansion is directly implemented in the inner \texttt{for c in range(4):} loop of our algorithm. The vector expression $t'_2 \boldsymbol{R}_{ab}^{(c)} + t'_3 \boldsymbol{D}_{ab}^{(c)}$ is handled by the \texttt{combined\_cross} variable, and the phase projection is finalized in \texttt{term\_NNN}. 

Finally, to ensure numerical stability and strict Hermiticity of the Hamiltonian, the total matrix is explicitly symmetrized via $H(\boldsymbol{k}) = [H(\boldsymbol{k}) + H^\dagger(\boldsymbol{k})]/2$ before diagonalization.

\section*{Supplementary Note 2: Mean-Field Decoupling of the On-Site Hubbard Interaction}

In this section, we provide the detailed derivation of the mean-field Hamiltonian for the on-site Coulomb interaction. The exact on-site Hubbard interaction is given by:
\begin{equation}
\mathcal{H}_{\text{on-site}} = U \sum_{a} \hat{n}_{a,\uparrow} \hat{n}_{a,\downarrow},
\end{equation}
where $\hat{n}_{a,\uparrow}$ and $\hat{n}_{a,\downarrow}$ are the particle number operators for the respective spin states at sublattice $a$. We can rewrite this interaction term using the total particle number operator $\hat{n}_a = \hat{n}_{a,\uparrow} + \hat{n}_{a,\downarrow}$ and the local spin magnetization operator along the quantization axis $\hat{M}_a = \hat{n}_{a,\uparrow} - \hat{n}_{a,\downarrow}$:
\begin{equation}
U \hat{n}_{a,\uparrow} \hat{n}_{a,\downarrow} = \frac{U}{4} \left[ (\hat{n}_a)^2 - (\hat{M}_a)^2 \right].
\end{equation}
To generalize this to an arbitrary spin direction, we define the vector magnetization operator as $\hat{\boldsymbol{m}}_a = c_a^\dagger \boldsymbol{\sigma} c_a$. Within the Hartree-Fock mean-field approximation, we expand the operator $\hat{\boldsymbol{m}}_a$ around its local expectation value. Let the expectation value be $\langle \hat{\boldsymbol{m}}_a \rangle = \boldsymbol{m}_a$, where $\boldsymbol{m}_a$ is the local magnetization vector (containing both magnitude and direction) at sublattice $a$. The operator can then be written as:
\begin{equation}
\hat{\boldsymbol{m}}_a = \boldsymbol{m}_a + \delta\hat{\boldsymbol{m}}_a,
\end{equation}
where $\delta\hat{\boldsymbol{m}}_a$ represents the small quantum fluctuation of the spin. Substituting this into the squared magnetization term and neglecting the second-order fluctuation term $(\delta\hat{\boldsymbol{m}}_a)^2$, we obtain:
\begin{align}
(\hat{\boldsymbol{m}}_a)^2 &= (\boldsymbol{m}_a + \delta\hat{\boldsymbol{m}}_a)^2 \nonumber \\
&\approx |\boldsymbol{m}_a|^2 + 2 \boldsymbol{m}_a \cdot \delta\hat{\boldsymbol{m}}_a \nonumber \\
&= |\boldsymbol{m}_a|^2 + 2 \boldsymbol{m}_a \cdot (\hat{\boldsymbol{m}}_a - \boldsymbol{m}_a) \nonumber \\
&= 2 \boldsymbol{m}_a \cdot \hat{\boldsymbol{m}}_a - |\boldsymbol{m}_a|^2.
\end{align}
Substituting Eq. (10) back into the generalized Hubbard interaction, we get:
\begin{equation}
\mathcal{H}_{\text{on-site}} \approx \frac{U}{4} \sum_a \left[ (\hat{n}_a)^2 + |\boldsymbol{m}_a|^2 - 2 \boldsymbol{m}_a \cdot \hat{\boldsymbol{m}}_a \right].
\end{equation}
Taking into account a strict half-filling condition, the total number of particles per site $\langle \hat{n}_a \rangle$ is constant. Consequently, the first two terms within the bracket simply contribute to a uniform scalar energy shift, which we can redefine as a constant $C$. Since a uniform global energy shift does not alter the topological band structure or the eigenstates, it can be safely dropped from the effective Bloch Hamiltonian. 

Mapping the many-body operator $\hat{\boldsymbol{m}}_a$ into the single-particle matrix representation ($\hat{\boldsymbol{m}}_a \rightarrow \boldsymbol{\sigma}$), the effective mean-field Hamiltonian for the general magnetic exchange interaction is derived as:
\begin{equation}
\mathcal{H}_{\text{on-site}}^{\text{MF}} = -\frac{U}{2} \sum_a \boldsymbol{m}_a \cdot \boldsymbol{\sigma} + const.
\end{equation}

For the specific numerical implementation of the canted magnetic state in our model, we assume that the magnitude of the magnetization is uniform across all four sublattices in the primitive cell, i.e., $|\boldsymbol{m}_a| = m$. By expressing $\boldsymbol{m}_a = m \hat{\boldsymbol{n}}_a(\theta)$, where $\hat{\boldsymbol{n}}_a(\theta)$ is the unit vector of direction, the Hamiltonian takes its final form:
\begin{equation}
\mathcal{H}_{\text{on-site}}^{\text{MF}} = -\frac{Um}{2} \sum_a \hat{\boldsymbol{n}}_a(\theta) \cdot \boldsymbol{\sigma}.
\end{equation}

\section*{Supplementary Note 3: Berry Curvature and Anomalous Transport Theory}
To understand profound impact of topological charge on the macroscopic transport properties, we consider the effective two-band Hamiltonian near a given Weyl node, which takes the minimal form $H(\boldsymbol{k}) = \pm v_F \boldsymbol{\sigma} \cdot \boldsymbol{k}$, where $v_F$ is the Fermi velocity and $\boldsymbol{\sigma}$ represents the Pauli matrices. From this Hamiltonian, one can derive the Berry curvature $\boldsymbol{\Omega}(\boldsymbol{k})$, defined as the curl of the Berry connection. Near the Weyl node, the Berry curvature exhibits a singular behavior mathematically isomorphic to the magnetic field generated by a point charge:
\begin{equation}
\boldsymbol{\Omega}(\boldsymbol{k}) = \pm \frac{\boldsymbol{k}}{2\vert{}\boldsymbol{k}\vert{}^3}
\end{equation}
Here, the $\pm$ sign dictates the topological charge (Chern number $C = \pm 1$) of the node, indicating whether it acts as a source or a sink of the Berry curvature flux~\cite{nagaosa2010anomalous}.
While the two-band effective model provides an intuitive picture of the monopole nature of Weyl nodes, in a realistic multi-band system like the pyrochlore lattice, the precise distribution of the Berry curvature must be evaluated across all relevant bands.
\\
To understand the transport properties driven by the Berry curvature, we consider the semiclassical equations of motion for an electron wave packet in a Bloch band. In the presence of an external electric field $\boldsymbol{E}$, the velocity of the electron is not only determined by the band dispersion but also acquires a transverse anomalous velocity term:
\begin{equation}
    \dot{\boldsymbol{r}} = \frac{1}{\hbar} \frac{\partial E_n(\boldsymbol{k})}{\partial \boldsymbol{k}} - \dot{\boldsymbol{k}} \times \boldsymbol{\Omega}_n(\boldsymbol{k})
\end{equation}
Since the rate of change of momentum is given by $\hbar \dot{\boldsymbol{k}} = -e\boldsymbol{E}$, the velocity equation can be rewritten as:
\begin{equation}
    \boldsymbol{v}_n(\boldsymbol{k}) = \boldsymbol{v}_{band} + \frac{e}{\hbar} \boldsymbol{E} \times \boldsymbol{\Omega}_n(\boldsymbol{k})
\end{equation}
This second term is the anomalous velocity, which is strictly perpendicular to both the applied electric field and the Berry curvature~\cite{xiao2010berry}. Near the Weyl nodes, where the Berry curvature $\boldsymbol{\Omega}_n(\boldsymbol{k})$ diverges, this anomalous velocity dominates the transverse electron transport, providing the microscopic origin of the intrinsic Anomalous Hall Effect without a net macroscopic magnetic field.
Using linear response theory, the $z$-component of the Berry curvature for the $n$-th energy band, $\Omega_{n}^{z}(\boldsymbol{k})$, is explicitly given by the Kubo formula formulation:
\begin{equation}
\Omega_{n}^{z}(\boldsymbol{k})=-2\hbar^2\sum_{m\neq n}\text{Im}[\frac{\langle n|v_x|m  \rangle \langle m|v_y|n \rangle}{(E_n-E_m)^2}]
\end{equation}
where $\vert{}n(\boldsymbol{k})\rangle$ and $E_n(\boldsymbol{k})$ are the Bloch wavefunction and the eigenvalue of the $n$-th band, respectively, and $v_i = \frac{1}{\hbar} \frac{\partial H}{\partial k_i}$ represents the velocity operator along the $i$-direction. This expression clearly demonstrates that the Berry curvature is drastically enhanced when two bands approach each other in energy ($E_n \rightarrow E_m$). Consequently, the dominant contributions to the total Berry curvature arise almost entirely from the immediate vicinity of the Weyl nodes where the band gap vanishes.

\section*{Supplementary Note 4: Exact position of weyl node in 3 dimension k-space}

\section*{1. Weyl Node Localization Algorithm}

Weyl nodes are points in the three-dimensional momentum space ($\boldsymbol{k}$-space) where two energy bands intersect linearly~\cite{wan2011topological,armitage2018weyl}. Our tight-binding model consists of 8 bands. Assuming a half-filling condition, we searched for the momentum $\boldsymbol{k}$ that minimizes the energy gap $\Delta E(\boldsymbol{k}) = E_5(\boldsymbol{k}) - E_4(\boldsymbol{k})$ between the highest occupied band (4th band) and the lowest unoccupied band (5th band). 

Furthermore, we have confirmed that throughout the full three-dimensional BZ--beyond the conventional one-dimensional high-symmetry lines--no additional degeneracies with adjacent subbands ($n = 3, 6$) occur on the local closed surfaces enclosing the Weyl nodes between the $n = 4$ and $n = 5$ bands. This absence of extraneous band touchings formally justifies restricting our low-energy topological charge analysis strictly to the crossings between the fourth and fifth bands~\cite{armitage2018weyl,xiao2010berry}. Consequently, the gapless points responsible for the topological phase are uniquely driven by the 4th and 5th bands. To prevent the optimization process from being trapped in local minima while searching for these exact points, we applied the following two-step algorithm.

\textbf{Global Grid Search and Clustering:} We discretized the Brillouin zone into a $30 \times 30 \times 30$ grid and calculated the energy gap at each point. After selecting candidates with $\Delta E(\boldsymbol{k}) < 0.1$, we removed duplicates by clustering points within a distance of $0.15$ from each other. This global search acts as a critical causal step, ensuring that the subsequent optimization algorithm initiates within the correct basin of attraction.

\textbf{Nelder-Mead Local Optimization:} We applied the derivative-free Nelder-Mead simplex algorithm to the clustered initial candidate vectors. The objective function was minimized with a strict tolerance of $10^{-9}$, ultimately pinpointing the precise three-dimensional coordinates of the Weyl nodes satisfying $\Delta E(\boldsymbol{k}) < 10^{-5}$.

\section*{2. Topological Chiral Charge Calculation}

In momentum space, a Weyl node acts as a magnetic monopole of Berry curvature~\cite{fang2003anomalous}. By integrating the Berry curvature flux over a closed surface enclosing the accurately located Weyl node, the topological chiral charge can be derived as a quantized integer~\cite{wan2011topological, xiao2010berry, fukui2005chern}.

\textbf{Abelian Berry Curvature:} We evaluated the velocity operator $v_i = \partial H / \partial k_i$ at an arbitrary $\boldsymbol{q}$ using the finite difference method (with a step size of $10^{-5}$). Considering the interband transitions between the occupied ($n \in \{1, 2, 3, 4\}$) and unoccupied ($m \in \{5, 6, 7, 8\}$) band spaces, the components of the Berry curvature vector $\bm{\Omega}(\boldsymbol{k}) = (\Omega_x, \Omega_y, \Omega_z)$ at a specific $\boldsymbol{k}$ point were calculated using the following Kubo formula form:
\begin{equation}
\Omega_\gamma(\boldsymbol{k}) = -2 \hbar^2 \sum_{n=1}^{4} \sum_{m=5}^{8} \text{Im} \left[ \frac{\langle \psi_n | v_\alpha | \psi_m \rangle \langle \psi_m | v_\beta | \psi_n \rangle}{(E_n - E_m)^2} \right]
\end{equation}
Here, $\alpha, \beta$, and $\gamma$ follow a cyclic permutation.

\textbf{Spherical Surface Integration:} We established an infinitesimal sphere with radius $R=0.01$ centered around the derived single Weyl node $\boldsymbol{k}_{\text{node}}$. The spherical surface was discretized by dividing the polar angle ($\theta$) into 40 segments and the azimuthal angle ($\phi$) into 80 segments, defining the infinitesimal area element as $d\boldsymbol{S} = R^2 \sin\theta d\theta d\phi \, \hat{\boldsymbol{r}}$. The final topological charge $C$ is derived as the total sum of the Berry curvature flux penetrating this closed surface:
\begin{equation}
C = \frac{1}{2\pi} \oint_{S} \bm{\Omega}(\boldsymbol{k}) \cdot d\boldsymbol{S}
\end{equation}

The result of this integration theoretically converges to an integer~\cite{xiao2010berry}. By rounding this calculated charge value to assign a spin characteristic of either $+1$ (Source) or $-1$ (Sink), we obtained a clear classification of the topological nature of each node.

The detailed numerical data for the Weyl node trajectories can be found in Supplementary Data 1 (weyl nodes trajectory.xlsx).
\end{document}